\documentclass[pra,superscriptaddress,twocolumn,floatfix]{revtex4-2}
\usepackage{graphicx}
\usepackage{physics}
\usepackage{ulem} % i inserted this to enable \sout command (stackexchange told me so).
\usepackage{color}
\definecolor{darkgreen}{rgb}{0,0.4,0} %% customized colors for the
\definecolor{darkblue}{rgb}{0,0,0.4}  %% for the
\definecolor{darkred}{rgb}{0.4,0,0}   %% hyper-referencing
\usepackage[colorlinks=true,%
    urlcolor=darkblue,citecolor=darkgreen,linkcolor=darkred]{hyperref}

\newcommand*{\Eq}[1]{\mbox{Eq.\ (\ref{eq:#1})}}
\newcommand*{\Fig}[1]{\mbox{Fig.\ \ref{fig:#1}}}

\begin{document}
\title{Experimental quantum state learning with pairs of photons}

% \date[]{\color{red}{Draft of 9 June 2026}\rule{0pt}{16pt}}

\date{\today}

\author{C.~Pria~Dobney}
\email{pria.dobney@mail.utoronto.ca}
\author{Johan~Henaff}
\author{Allen~Kasum}
\author{Rui~Jie~Tang}
\author{Haru~Mukumoto}

\affiliation{Department of Physics and %
  Centre for Quantum Information \& Quantum Control, %
  University of Toronto, Canada}

\author{Mark~Hillery}
\affiliation{Physics Program, %
  Graduate Center of the City University of New York, USA}

\author{Berthold-Georg~Englert}
\affiliation{School of Physics, %
  Beijing Institute of Technology, China}
\affiliation{Department of Physics, %
  National University of Singapore, Singapore}

\author{Aephraim~Steinberg}
% \email{steinberg@physics.utoronto.ca}
\affiliation{Department of Physics and %
  Centre for Quantum Information \& Quantum Control, %
  University of Toronto, Canada}
% \affiliation{Canadian Institute for Advanced Research, Toronto, Canada}

\begin{abstract}
Tomography allows one to estimate the density matrix describing the state an ensemble of quantum systems are prepared in (for example, polarization tomography determines the polarization state of a beam of identically prepared photons).
In general, it is not possible to uniquely decompose the density matrix into
its pure state components.
Agarwal \textit{et al.} proposed a protocol which, for a mixture composed of any two pure states of a qubit (with arbitrary probabilities), allows an observer to infer not only the density matrix but the identity of those specific pure states and their weights – the additional requirement being that the qubits arrive in pairs, where both qubits in each pair are in the same state.
We experimentally demonstrate this learning-from-pairs concept using photons
in the polarization degree of freedom. 
We use tomography to measure a sequence of single photons and make use of
their time-of-arrival information to `pair up' the photons after the
measurement. 
From here we are able to infer the photons' polarization states and their
respective probabilities, and we demonstrate this for various different
choices of polarization states and ratios. 
Finally, we investigate our ability to discriminate between two equal mixtures
of distinct pairs of orthogonal polarization states.  
We find that on the order of $\sim 10^4$ photons is typically enough to
achieve tomography fidelities of approximately 99.99\%. This is sufficient to discriminate between two different preparations of the same mixed state, differing by angles of less than $5^{\circ}$ between the pure states used in the two preparations.

\end{abstract}

\maketitle

\section{Introduction}
Quantum state learning, or estimation, is fundamental in quantum information
processing and communications and a central concept in quantum measurement
theory.  
Information is encoded into the degrees of freedom of a quantum system and it
is crucial to be able to reliably and repeatedly prepare and characterize such
states in order to determine the information encoded. 
Knowing the quantum state of a system allows one to make predictions about the future of that system \cite{hradil1997quantum, helstrom1969quantum}. 
Methods such as quantum state tomography are often used to infer the quantum
state (represented, for example, by a density matrix or a Wigner function \cite{wigner1932quantum}) from a set of data and measurements
performed on a large number of identically-prepared copies of the state under
test. 
By now, the field of tomography is well-developed: a non-exhaustive list of
different approaches include adaptive tomography \cite{mahler2013adaptive,
  adaptiveBayesian}, compressive tomography \cite{garberoglio2025compressive},
using mutually unbiased bases as the measurement projections
\cite{wootters1989mub, adamson2010mub}, and shadow tomography
\cite{aaronson2018shadow, huang2020shadow}. 

%move+merge para4-5 ("in this work...")
In this work, we consider a scenario where Alice wants to send a stream of
single qubits to Bob (in our case, the qubits are carried by single photons;
information is encoded in the polarization degree of freedom). 
Alice promises Bob that the photons are prepared in either one of two distinct
polarization states, which we call $\ket{\psi_0}$ and $\ket{\psi_1}$, and the respective probability of each state arising is either $p_0$ or
$p_1$. 
This sequence of photons could encode some useful and potentially private
information as in applications such as quantum key distribution
\cite{hklo2016practical}.  
Bob measures the sequence of single photons by performing single-photon
polarization tomography. 
He obtains the single-qubit density matrix of the ensemble,
\begin{equation}\label{eq:singlerho}
  \rho_{\mathrm{single}}
  = p_0 \ket{\psi_0}\!\bra{\psi_0} + p_1 \ket{\psi_1}\!\bra{\psi_1}\,.
\end{equation}
Bob's task is to learn the quantum states $\ket{\psi_0}$ and $\ket{\psi_1}$
and their probabilities $p_0$ and $p_1$ using only the data he measures. 
However, standard single-particle tomography is not sufficient to identify the specific states: it is well-known that there is in general no unique decomposition of a density matrix into pure states.

\begin{figure}
    \centering
    \includegraphics[width=0.95\linewidth]{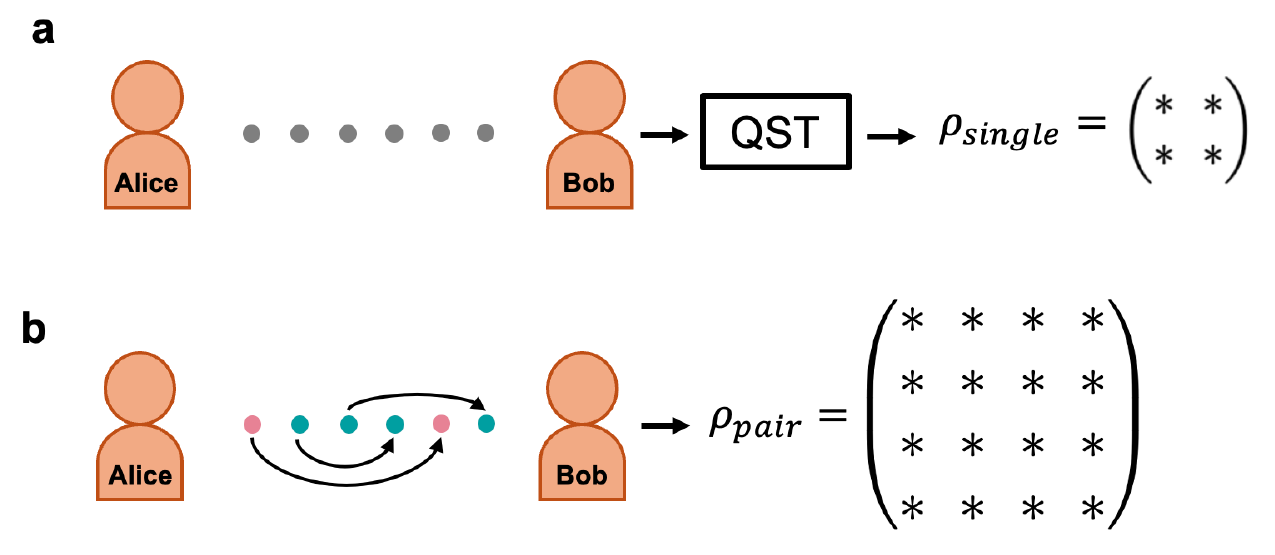}
    \caption{\label{fig:schem}%
      Schematic showing the experimental protocol.
      (a) Alice sends a stream of single photons to Bob.
      Bob performs quantum state tomography (QST) on the sequence and obtains
      the single-qubit density matrix, $\rho_{\mathrm{single}}$ of
      \Eq{singlerho}.
      (b) After the measurement, Alice tells Bob which photons are in a
      `pair'.
      From this `paired data', Bob obtains the two-qubit density matrix
      $\rho_{\mathrm{pair}}$ of \Eq{2rho}, which can be decomposed uniquely
      into pure product states and their respective probabilities.
      The pairs needn't be nearest neighbours (in time); any choice of pairing identification suffices.}
\end{figure}

Agarwal \textit{et al.} \cite{statelearning} show that if more information is provided in the form of an
extra copy of each state $\ket{\psi_{0}}$ or $\ket{\psi_{1}}$, then the relevant density matrix is the \textit{two}-qubit density matrix
\begin{equation}\label{eq:2rho}
  \rho_{\mathrm{pair}} = p_{0}\ket{\psi_{0}\psi_{0}}\!\bra{\psi_{0}\psi_{0}} + p_{1}\ket{\psi_{1}\psi_{1}}\!\bra{\psi_{1}\psi_{1}},
\end{equation}
which \textit{does} have a unique decomposition into two pure product states, and Bob can thus determine $\ket{\psi_{0}}$, $\ket{\psi_{1}}$, $p_{0}$ and $p_{1}$.
Here, we report on an experimental demonstration of this protocol.
Interestingly, this can be achieved regardless of whether Bob is given this
information before he measures the photons (i.e. Alice sends him pairs of
photons) or afterwards (Alice sends him single photons, then tells Bob which
two photons should be paired together); see Fig.~\ref{fig:schem} for an illustration of the latter approach.
This `learning from pairs' concept -- specifically the use of additional \textit{copies} of the states of interest -- is reminiscent of the fact that from single-particle measurements, one can obtain at most only the density matrix $\rho$, but with multi-particle measurements, it is possible to learn other functions that may be non-linear in $\rho$, which are inaccessible when one is limited to only single-particle measurements \cite{brun2004measuring, adamson2007preparation}.
% \sout{In this paper, we report on an experimental demonstration of this scenario,
% based on the theory by Agarwal \textit{et al.} \cite{statelearning}. 
% Here, the authors state that if more information is provided in the form of an
% extra copy of each state $\ket{\psi_{0}}$ or $\ket{\psi_{1}}$ then the density
% matrix in \Eq{singlerho} can be written as a \textit{two-}qubit density
% matrix,}
% \sout{\begin{equation}\label{eq:2rho}
%   \rho_{\mathrm{pair}} = p_{0}\ket{\psi_{0}\psi_{0}}\!\bra{\psi_{0}\psi_{0}} + p_{1}\ket{\psi_{1}\psi_{1}}\!\bra{\psi_{1}\psi_{1}}\,.
% \end{equation}}
% \sout{This has a unique decomposition into two pure product states.
% Bob can thus determine $\ket{\psi_{0}}$, $\ket{\psi_{1}}$, $p_{0}$ and $p_{1}$.
% Interestingly, this can be achieved regardless of whether Bob is given this
% information before he measures the photons (i.e. Alice sends him pairs of
% photons) or afterwards (Alice sends him single photons, then tells Bob which
% two photons should be paired together). }

Since no finite number of copies used in tomography can identify the `true' state with certainty, one obtains an estimate of the state. 
There will naturally be some inaccuracy between this estimation and the true
state. 
One can ask how `close' the estimated state is to the true state:
common ways of quantifying this distance include the fidelity (or infidelity)
or the trace distance \cite{wootters1981statistical, helstrom1969quantum,
  jozsa1994fidelity}.  
One may also quantify the imprecision of the estimated state by finding a
credible region which contains the true state with a high probability
\cite{evans2006surprise,shang2013regions,englert2025lqse}. 
% \textcolor{magenta}{wondering if it is worth mentioning that the various criteria with fidelity or trace distance rely on knowing the "true" state whereas the credible regions (and all that) only use the actual data acquired by Bob. You see, in the setting of Figure 1, Bob cannot calculate the fidelity or the trace distance, but he can identify the credible regions. }

Quantum state discrimination \cite{herzog2004distinguishing} explores how
confidently we can distinguish one quantum state from at least one other.   
The goal is usually to determine the actual state of a quantum system which is
prepared in a certain but unknown state from a set of possible states with
some prior probability.  
However, if the set of possible states are not mutually orthogonal, it becomes
harder to unambiguously distinguish one from the other. 
This has been extensively explored for pure states
\cite{ivanovic1987differentiate, dieks1988overlap, peres1988differentiate} and
experimentally demonstrated \cite{huttner1996unambiguous}.  
A similar problem is the one of quantum state comparison
\cite{barnett2003comparison}, which asks if we can establish whether or not
two quantum systems have been prepared in the same state.  
Discrimination of mixed quantum states has also been studied: unambiguous discrimination (where the probability of error is zero) is
possible between two orthogonal mixed states \cite{rudolph2003unambiguous}.  
In general, for the case of discriminating between non-orthogonal mixtures,
one of two main approaches is followed: optimal unambiguous discrimination
\cite{herzog2005optimum, clarke2001experimental}, where the probability of an
inconclusive result is minimised, or minimum-error discrimination
\cite{herzog2002minimum}, which allows for ambiguity in the measurement
outcome but with the smallest possible error.  
The problem of quantum state filtering \cite{sun2002filter}, where one wishes
to discriminate between two subsets of a set of non-orthogonal states, can be
cast as the case of discrimination between a pure state and a mixed state, and
has been experimentally demonstrated in an optical interferometer system
\cite{mohseni2004optical}, where the authors also explored the differences
between using generalised measurements and projective
measurements~\cite{herzog2004distinguishing}.

In our experiment, we make use of the time-of-arrival information for each detected photon, recorded
during single-photon tomography measurements, and group each single photon
into a pair with another photon in the ensemble, as depicted in \Fig{schem}. 
The pairs are then used to obtain the two-photon density matrix
$\rho_{\mathrm{pair}}$. 
We investigate our ability to distinguish between two states $\ket{\psi_0}$
and $\ket{\psi_1}$ in the statistical mixture forming $\rho_{\mathrm{single}}$
as these two states get closer to each other on the Poincar\'e sphere. 
Following this, we distinguish between two mixed states represented by the
\emph{same} single-qubit density matrix $\rho_{\mathrm{single}}$, but which
are composed of \textit{different} $\ket{\psi_0}$ and $\ket{\psi_1}$, so that their two-qubit density matrices
 $\rho_{\mathrm{pair}}$ are \emph{not} the same.

\section{Learning from pairs}
\begin{figure}
    \centering
    \includegraphics[width=0.95\linewidth]{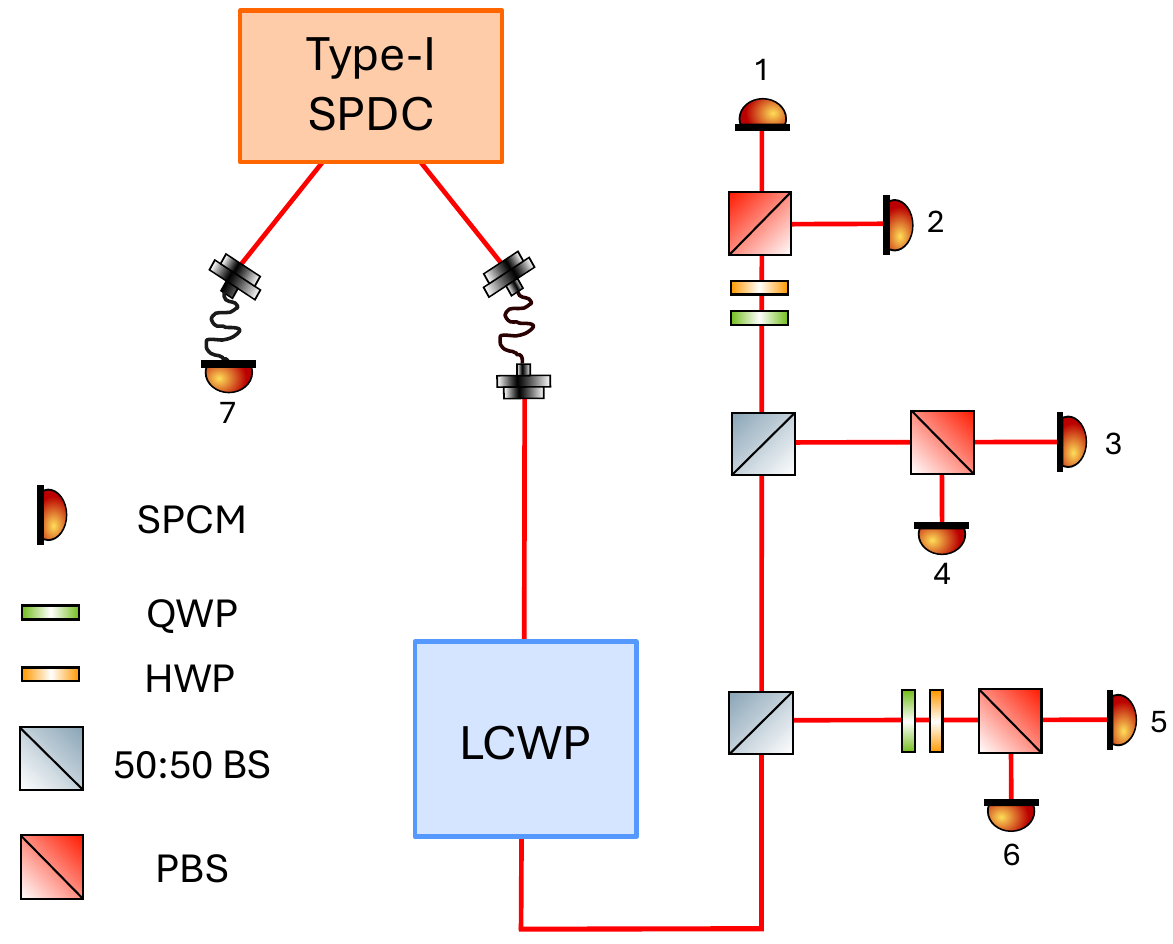}
  \caption{\label{fig:statelearn}%
    Schematic of the experimental setup.
    A type-I, degenerate, non-collinear spontaneous parametric down-conversion (SPDC) process where
    on average 3500 pairs of 808~nm photons are detected per second.
    One photon is immediately detected whilst the other is sent through the
    experiment.
    These heralded single photons pass through a liquid-crystal wave plate
    (LCWP) which rotates the polarization of the incident photons by an amount
    dependent on the voltage applied to the crystal in order to generate a
    statistical mixture of polarization states $\ket{\psi_0}$ and
    $\ket{\psi_1}$ as in \Eq{singlerho}.
    The ensemble of single photons is measured at six different output ports
    of a polarization tomography, where each of the remaining six
    single-photon counting modules (SPCMs) measures in one of the six
    polarization bases.
    Detectors 1 and 2 measure photons in the right-handed (R) and left-handed
    circular (L) basis (R/L);
    detectors 3 and 4 measure in the horizontal (H) and vertical (V) basis;
    and detectors 5 and 6 measure in the diagonal (D) and anti-diagonal (A)
    basis. 
    QWP: quarter-wave plate; HWP: half-wave plate;
    BS: beamsplitter; PBS: polarizing beamsplitter. }
\end{figure}

Our experiment (see Fig.~\ref{fig:statelearn}) begins with a degenerate
spontaneous parametric down-conversion (SPDC) source to generate pairs of
single photons at 808~nm.  
A $\beta-$BBO (beta-barium borate) crystal used for second harmonic generation
(SHG) is pumped with 808~nm light from a Ti:Sapphire pulsed laser (Coherent
Chameleon Ultra II), with a 140~fs pulse duration and 80~MHz repetition rate \footnote{The use of pulsed laser light as the pump is not strictly necessary; our experiment only requires timing information for the purpose of the pairing process, for which the heralding should suffice so long as the photon detection rate is not too high.}.   
The 404~nm light produced from SHG is used to pump a second $\beta-$BBO
crystal, which is 2~mm thick and cut at $29.3^{\circ}$ for type-I, non-collinear SPDC, at a time-averaged power of 12~mW. 
The 404~nm light is focused to a waist of $\sim 70~\mu$m. 
The down-converted photons are produced at an emission angle of $3.2^{\circ}$ and after passing through 10~nm bandpass filters (centre wavelength of 810~nm) are collected using single-mode fibres, with a waist of $\sim 50~\mu$m.

Roughly 3500 coincidences (photon pairs) per second are detected from the source, using single-photon counting modules (SPCMs, Perkin-Elmer SPCM-AQRH series) with a 4~ns coincidence window. 
Approximately 30000 single photons per second are measured at each detector.
The SPCMs used in the experiment have dark count rates ranging from approximately 1 to 4~kHz, and the rate of accidentals is less than 5 per second measured directly from the source and negligible in the tomography stage of the experiment.

The experiment proceeds as follows. After leaving the source, the idler photon is detected immediately (detector
7), acting as a trigger.  
The signal photon passes through a polarizing beamsplitter (PBS) in order to
prepare it in the state $\ket{\mathrm{H}}$. 
A liquid crystal wave plate (LCWP) at $45^{\circ}$ applies a polarization
rotation to the incident light, where the retardance is varied depending on
the applied voltage to the LCWP (between 0 and $20~\mathrm{V}$). 
The LCWP switches the polarization rotation applied to the signal photons
between two voltage settings to generate the mixture of $\ket{\psi_0}$ and
$\ket{\psi_1}$. 
The proportion of the total time of the acquisition that the LCWP prepares
either $\ket{\psi_0}$ or $\ket{\psi_1}$ determines the probabilities $p_0$ and
$p_1$, respectively. 

The signal photons are detected using a single-photon polarization tomography
apparatus depicted on the right-hand side of \Fig{statelearn}. 
Each of the three branches of the tomography stage corresponds to a projection
onto each of the three polarization bases, namely R/L, H/V, and D/A. 
SPCMs and a time-tagger (Swabian Instruments) are used to measure coincidence events
between the idler photons across the six tomography measurement outcomes
whilst recording the corresponding times-of-arrivals of the photons. 
From these rates, we can obtain the single-photon density matrix
$\rho_{\mathrm{single}}$ of \Eq{singlerho} for the ensemble.

% transition paragraph
The photon rate data from each of the six polarization measurements and their
timestamps were then used to produce `paired data' from which we obtain the
two-photon density matrix $\rho_{\mathrm{pair}}$ of \Eq{2rho}.
Following the procedure outlined in Ref.~\cite{statelearning},
$\rho_{\mathrm{pair}}$ can be decomposed into the pure state vectors we wish to obtain.
Using our setup, we experimentally probe various ways that information can be encoded in different mixtures of two pure polarization states. 
First, we look at a single mixture of two pure states, and find the limit on how close these two states can be before we cannot discriminate between them.
We then consider multiple different preparations of the same mixture and our ability to distinguish between two such preparations.  \\

% \section{Results}
%
\subsection{Distinguishing between pure states in a mixture}\label{sec:pop}
\begin{figure*}
    \centering
    \includegraphics[width=0.96\textwidth]{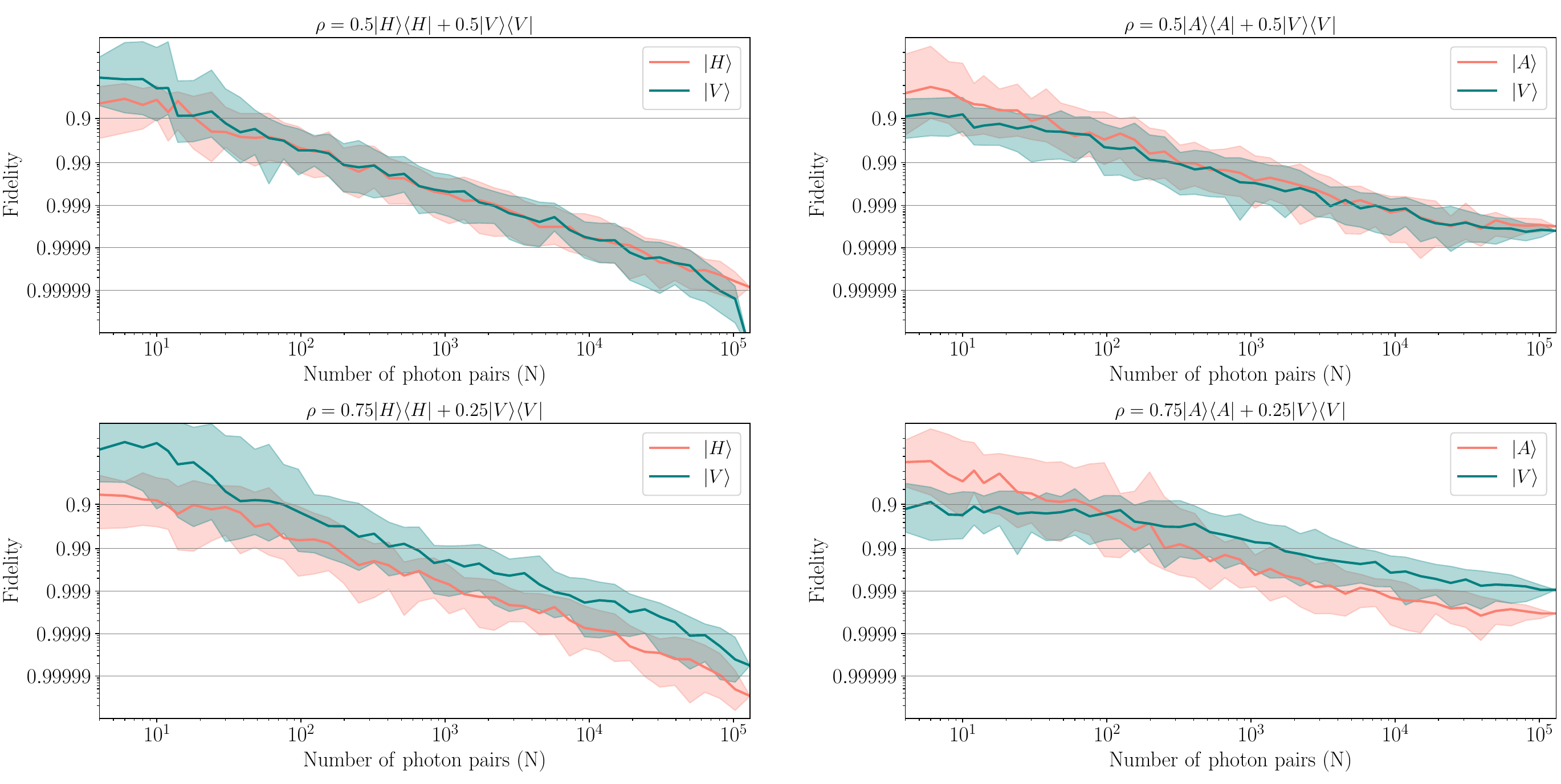}
    \caption{\label{fig:pop_hvva}%
      Averaged fidelities vs. number of photon pairs of each of the states
      (pink curves: $\ket{\psi_0}$; teal curves: $\ket{\psi_1}$) determined
      from the decomposition of the paired density matrix $\rho'$ in
      \Eq{2rho}.
      (a) and (b): $\ket{\psi_0}$ and $\ket{\psi_1}$  are orthogonal to each
      other but appear (a) with equal probability (${p_{0}=p_{1}=0.5}$) and
      (b) with unequal probability (${p_{0}=0.75}$, ${p_{1}=0.25}$). 
      (c) and (d): $\ket{\psi_0}$ and $\ket{\psi_1}$ are \emph{not} orthogonal
      to each other but appear (c) with equal probability
      (${p_{0}=p_{1}=0.5}$) and (d) with unequal probability (${p_{0}=0.75}$,
      ${p_{1}=0.25}$). 
      The solid lines are the median values after 50 repeats of the fidelity calculation.
      Since the fidelity is not a linear metric and our data are not
      Gaussian-distributed, we calculate the bounds of the shaded regions on
      all four plots (a)-(d) from the 16th and 84th percentiles (found by linearly interpolating between two consecutive points).
      In (a) and (b), where $\ket{\psi_0}$ and $\ket{\psi_1}$ are orthogonal to each other, we observe that the curves scale approximately as $1-F \propto 1/N$, as expected theoretically \cite{mahler2013adaptive}. The curves in plots (c) and (d) (where $\ket{\psi_0}$ and $\ket{\psi_1}$ are not orthogonal) exhibit regions where the $1/N$ scaling is roughly followed, but appear to plateau beyond ${N\sim10^{4}}$ measured photon pairs.
      We notice that in (d), the two curves cross.
      When compared with (c), it can be seen that the pink curves (for
      $\ket{\mathrm{A}}$) follow the same trend in both plots, whereas the
      teal curve (for $\ket{\mathrm{V}}$) in (d) does not reach the same
      fidelity as in (c) even after $\sim10^{5}$ photon pairs are detected.
      This is because the proportion of $\ket{\mathrm{V}}$ is lower than the
      proportion of $\ket{\mathrm{A}}$ in the mixture used in (d) (and similarly in (b), where $\ket{H}$ appears with higher probability than $\ket{V}$).
      % \query{The labeling and text in the plots is rather small. Can it be larger?}
      }    
\end{figure*}

We characterize the performance of our protocol by calculating the fidelity of the state vectors thus obtained with the `true’ states we attempted to create in our state preparation.
The states $\ket{\psi_{0}}$ and $\ket{\psi_{1}}$ obtained from the paired
density matrix $\rho_{\mathrm{pair}}$ are used to construct density matrix
representations of these states, ${\rho_{0} = \ket{\psi_0}\!\bra{\psi_0}}$ and
${\rho_{1} = \ket{\psi_1}\!\bra{\psi_1}}$. 
From the overlap between each density matrix and the expected state determined from single-photon tomography measurements ($\rho_{0}^{\mathrm{exp}}$ or $\rho_{1}^{\mathrm{exp}}$,
respectively), which were performed separately, we estimate the fidelity as $ F = \mathrm{Tr}\left(\rho_{j} \rho_{j}^{\mathrm{exp}} \right) $ where $j = 0,1$.
For different selections of $\ket{\psi_{0}}$, $\ket{\psi_{1}}$, $p_{0}$ and
$p_{1}$, we calculate the fidelities $F$ as a function of the number $N$ of detected photon
pairs (see \Fig{pop_hvva}).

Plots (a) and (b) in \Fig{pop_hvva} show the fidelity vs.\ the number of
detected pairs when $\ket{\psi_0}$ (pink curves) and $\ket{\psi_1}$ (teal
curves) are orthogonal to each other (namely,
${\ket{\psi_0}=\ket{\mathrm{V}}}$ is prepared vertically polarized and
${\ket{\psi_1}=\ket{\mathrm{H}}}$ is horizontally polarized).  
In (a), the two states appear with equal probability (${p_{0}=p_{1}=0.5}$); in
(b), ${p_{0}=0.75}$, ${p_{1}=0.25}$. 
The solid lines for $\ket{\psi_0}$ and $\ket{\psi_1}$ have been averaged over
50 repeats of the pair-decomposition calculation and the median is plotted.  
We find that from ${N \sim 10^{4}}$ pairs of photons, we are able to learn the
two states with fidelities of 99.99\%, and in (b) observe that the state with
the higher probability reaches this sooner. 

Plots (c) and (d) again show the fidelity vs.\ the number of detected pairs
but for the case when $\ket{\psi_0}$ and $\ket{\psi_1}$ are not orthogonal to
each other. 
Here, they are separated by $\pi/2$ radians on the Poincaré sphere
(${\ket{\psi_0}=\ket{\textrm{V}}}$ is vertically polarized and
${\ket{\psi_1}=\ket{\mathrm{A}}}$ is anti-diagonally polarized).
We see that, as is expected, more pairs of photons are required to reach the
same level of fidelity as achieved for the case where $\ket{\psi_0}$ and
$\ket{\psi_1}$ are orthogonal to each other \cite{statelearning}. 
We observe that all the fidelity curves follow the general trend of
${1-F\propto 1/N}$, in agreement with the theoretical prediction \cite{mahler2013adaptive}.

\begin{figure}[h]
    \centering
    \includegraphics[width=0.96\linewidth]{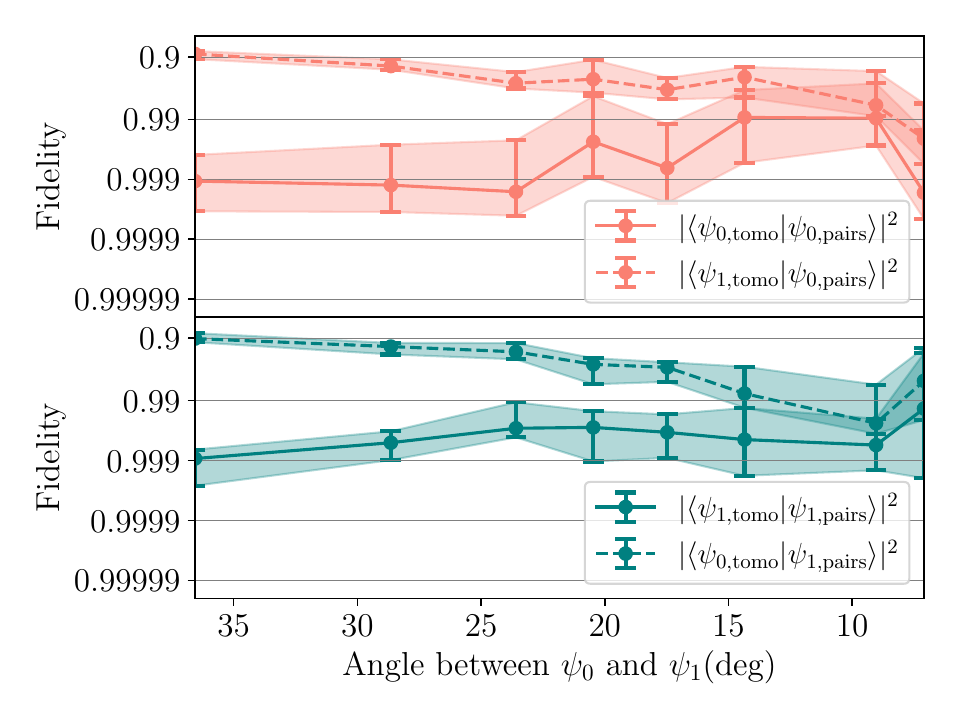}
    \caption{\label{fig:anglevpair}%
      Plots of the fidelity between the two states in the ensemble
      ($\ket{\psi_0}$ and $\ket{\psi_1}$) against the angle between the two
      state vectors. 
      Points connected by dashed lines are the fidelity between state
      $\ket{\psi_{\mathrm{0,pairs}}}$ ($\ket{\psi_{\mathrm{1,pairs}}}$)
      calculated from the pair-state density matrix with
      $\ket{\psi_{\mathrm{1,tomo}}}$ ($\ket{\psi_{\mathrm{0,tomo}}}$) measured
      from single-qubit tomography. 
      In the upper (lower) plot, points connected by solid lines are the
      fidelity between $\ket{\psi_{\mathrm{0,pairs}}}$
      ($\ket{\psi_{\mathrm{1,pairs}}}$) from the pair-state density matrix and
      $\ket{\psi_{\mathrm{0,tomo}}}$ ($\ket{\psi_{\mathrm{1,tomo}}}$) measured
      from single-qubit tomography. 
      As the two states get closer to each other on the Poincar\'e  sphere, it
      becomes harder to distinguish the two, and hence we see the two curves
      overlapping for separations of around $\sim 15^{\circ}$ and less. 
    Error bars are calculated from repeating the experiment 10 times, each
    with $N \sim 7 \times 10^4 $ pairs of photons, and the data points are the
    median values. }     
\end{figure}

As the angle between the two state vectors $\ket{\psi_0}$ and $\ket{\psi_1}$
decreases, it becomes harder to distinguish the two states. 
We are interested in how close the two states in a mixture can be to one
another before we are unable to tell them apart. 
In \Fig{anglevpair}, we plot the fidelity between $\ket{\psi_0}$ and
$\ket{\psi_1}$ as the angle between the two becomes smaller.
% \query{Or ``becomes decreasingly smaller'' or just ``becomes smaller''
%   or ``decreases''?} 
In the upper (lower) plot, points connected by dashed lines are the fidelity
between state $\ket{\psi_{\mathrm{0,pairs}}}$
($\ket{\psi_{\mathrm{1,pairs}}}$) calculated from the pair-state density
matrix with $\ket{\psi_{\mathrm{1,tomo}}}$ ($\ket{\psi_{\mathrm{0,tomo}}}$)
expected from single-qubit tomography;
points connected by solid lines are the fidelity between
$\ket{\psi_{\mathrm{0,pairs}}}$ ($\ket{\psi_{\mathrm{1,pairs}}}$) from the
pair-state density matrix and $\ket{\psi_{\mathrm{0,tomo}}}$
($\ket{\psi_{\mathrm{1,tomo}}}$) expected from single-qubit tomography. 
Single-qubit tomography measurements were performed in a separate data run
(as described in \Fig{pop_hvva}) in order to determine
$\ket{\psi_{\mathrm{1,tomo}}}$ and $\ket{\psi_{\mathrm{0,tomo}}}$. 
In both plots the number of photon pairs $N \sim 7 \times 10^{4}$ is fixed,
and error bars are derived from repeating the process 10~times, and the data
points show the median value. 
The error bars depend on the numerical optimization method used to calculate
the states from the density matrices returned from tomography.
We found that overall, using a differential evolution algorithm for this
process resulted in smaller deviations per point compared with other
algorithms such as regular maximum likelihood.  
We see that, at angles starting around $15^{\circ}$ and lower, the two states
cannot be distinguished.
At small angles it becomes harder to confidently tell the two states
$\ket{\psi_0}$ and $\ket{\psi_1}$ in the ensemble apart and thus the two
curves on each plot start to overlap. 
This shows that we cannot employ mixtures of two pure states that are separated by less than $15^{\circ}$.

\subsection{Distinguishing between different mixtures of orthogonal states}
\label{sec:multpairs}
A greater number of distinct mixtures can be used to encode information if the
separation between two mixed states can be decreased, as long as the two preparations remain (pair-)distinguishable from one another.
Here, we investigate how well we can discriminate between different mixtures
of orthogonal states as the separation between the two mixed states
decreases. 
Alice sends Bob various (equally-weighted) mixtures of two photons in
orthogonal polarization states.  
Bob needs to be able to distinguish between two such mixtures, $\rho$ and
$\rho'$, as the angle between $\ket{\psi_0}$ and $\ket{\psi_0'}$ is decreased,
where the single-photon density matrices $\rho$ and $\rho'$ can be written as 
\begin{align}
  \rho = \frac{1}{2} \Bigl( \ket{\psi_0}\!\bra{\psi_0}
  + \ket{\psi_1}\!\bra{\psi_1} \Bigr); \\
  \rho' = \frac{1}{2} \Bigl( \ket{\psi_0'}\!\bra{\psi_0'}
  + \ket{\psi_1'}\!\bra{\psi_1'} \Bigr)\,.
\end{align} 

\begin{figure*}
    \centering
    \includegraphics[width=0.98\textwidth]{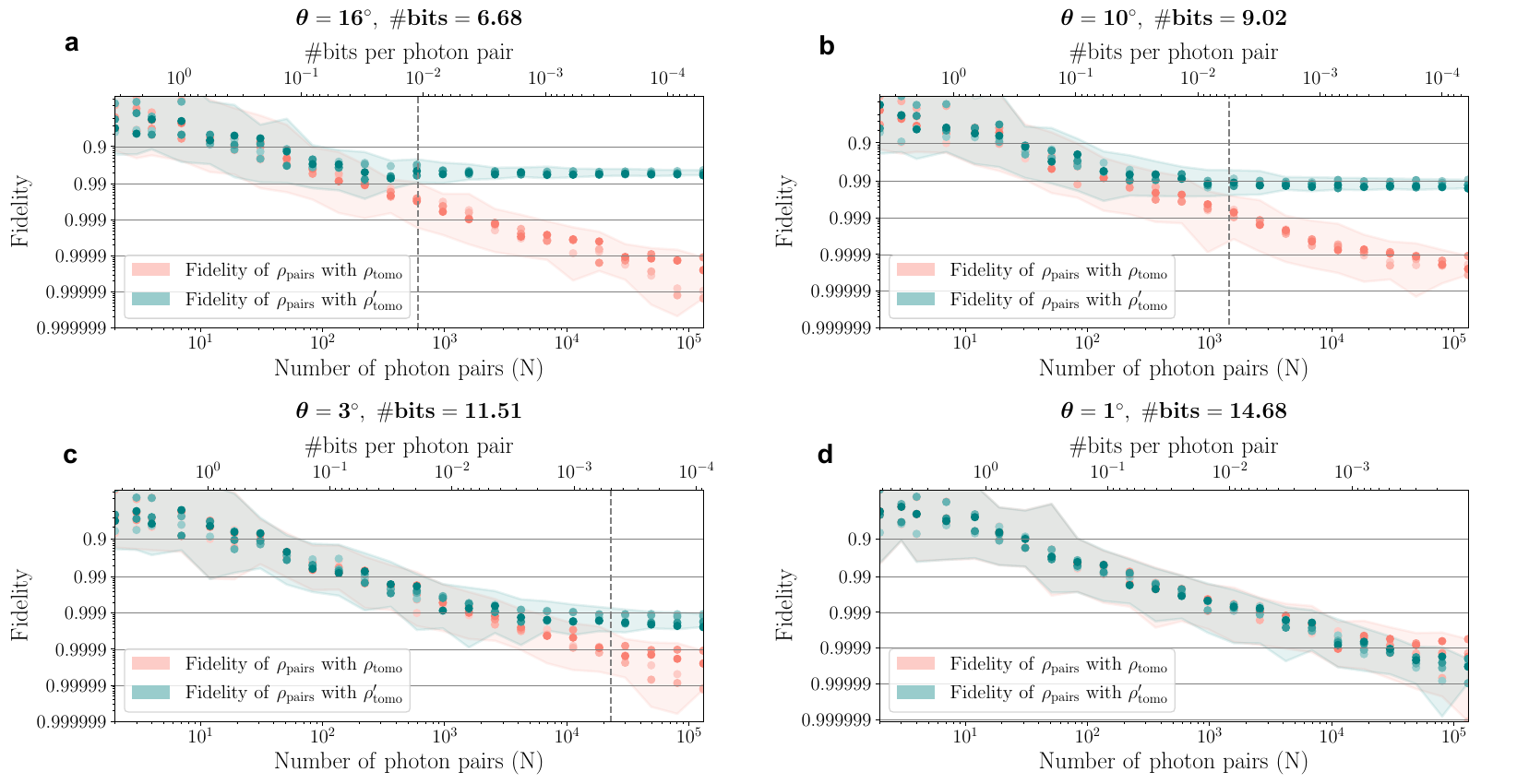}
    \caption{\label{fig:multpairsplots}%
        Fidelity calculated between the pure states in one of two distinct
        mixtures as a way to compare the two mixed states, in terms of the
        number of photon pairs $N$ as the distance (angle $\theta$, in degrees) between
        the two mixtures decreases.
        Overall, there are two main collections of points forming curves on these four plots.
        For each case, the fidelity is calculated between the states returned from the pair-state density matrix calculation and the expected state from single-qubit tomography. The pink points are the fidelities between
        $\ket{\psi_{\mathrm{0, pairs}}}$ ($\ket{\psi_{\mathrm{1, pairs}}}$) from the pair-state density matrix and
        $\ket{\psi_{\mathrm{0, tomo}}}$ ($\ket{\psi_{\mathrm{1, tomo}}}$) from single-photon tomography measurements, as well
        as for between $\ket{\psi_{\mathrm{0, pairs}}'}$ ($\ket{\psi_{\mathrm{1, pairs}}'}$) and $\ket{\psi_{\mathrm{0, tomo}}'}$ ($\ket{\psi_{\mathrm{1, tomo}}'}$).
    The teal points are calculated similarly, however this time the
    fidelity is calculated between $\ket{\psi_{\mathrm{0, pairs}}}$ ($\ket{\psi_{\mathrm{1, pairs}}}$) and
    $\ket{\psi_{\mathrm{0, tomo}}'}$ ($\ket{\psi_{\mathrm{1, tomo}}'}$), as well as between $\ket{\psi_{\mathrm{0, pairs}}'}$ ($\ket{\psi_{\mathrm{1, pairs}}'}$) and
    $\ket{\psi_{\mathrm{0, tomo}}}$ ($\ket{\psi_{\mathrm{1, tomo}}}$)
    As the number of photon pairs detected increases, the point where the
    shaded regions around the two sets of curves (pink and teal) no longer
    overlap is the point at which we can distinguish the two mixtures (marked by vertical dashed lines).
    As the two states get closer, more photons are needed. 
    All of the curves generally have a region where the slope goes as $1/N$, as
    expected. 
    The number of bits shown is calculated by $\log_2{(8/\theta^{2})}$ for
    each value of $\theta$ in (a), (b), (c), and (d).} 
    \end{figure*}

In \Fig{multpairsplots}, we calculate the fidelity between the state
calculated from the coincidence density matrix with the expected state from
single-photon tomography, and investigate how well we can differentiate
between the two states as the separation is decreased, from $16^{\circ}$ in
(a) to $1^{\circ}$ in (d). 
In all plots (a) to (d), the pink points show the fidelity between state
$\ket{\psi_{\mathrm{0, pairs}}}$ ($\ket{\psi_{\mathrm{1, pairs}}}$) from the paired density matrix with $\ket{\psi_{\mathrm{0, tomo}}}$ ($\ket{\psi_{\mathrm{1, tomo}}}$), the expected state measured from single-qubit
tomography, and for the fidelity between $\ket{\psi_{\mathrm{0, pairs}}'}$ ($\ket{\psi_{\mathrm{1, pairs}}'}$) and $\ket{\psi_{\mathrm{0, tomo}}'}$ ($\ket{\psi_{\mathrm{1, tomo}}'}$). 
The teal points show the fidelity between state $\ket{\psi_{\mathrm{0, pairs}}}$ ($\ket{\psi_{\mathrm{1, pairs}}}$) and
$\ket{\psi_{\mathrm{0, tomo}}'}$ ($\ket{\psi_{\mathrm{1, tomo}}'}$), as well as between $\ket{\psi_{\mathrm{0, pairs}}'}$ ($\ket{\psi_{\mathrm{1, pairs}}'}$) and
$\ket{\psi_{\mathrm{0, tomo}}}$ ($\ket{\psi_{\mathrm{1, tomo}}}$).

As in \mbox{Figs.\ \ref{fig:pop_hvva}} and \ref{fig:anglevpair}, separate
single-qubit states were prepared and measured to obtain the states
${\rho_{\mathrm{tomo}}=\ket{\psi_{i,\mathrm{tomo}}}\!\bra{\psi_{i,\mathrm{tomo}}}}$
and
${\rho'_{\mathrm{tomo}}=\ket{\psi'_{i,\mathrm{tomo}}}\!\bra{\psi'_{i,\mathrm{tomo}}}}$
for ${i \in \{0,1\}}$. 
We observe the same general trend in the (in)fidelity as in
\Fig{pop_hvva}: namely, there are regions where the slope is roughly $1/N$. 
 
The point (in terms of $N$) at which there is no longer an overlap between the
teal and the pink is the number of photon pairs required to distinguish
between the two mixtures (indicated by the vertical dashed lines).
When the two mixtures are more dissimilar to each other (larger angle
separation, seen in plots (a) and (b)), the teal curves appear to plateau
faster to a lower fidelity than the pink curves, and fewer photon pairs are
required to see this than in plots (c) and (d).  

For each case presented in \Fig{multpairsplots}, we calculate how many
different distinct mixtures can be made, each separated by the given angle.
From this, we can determine the maximum number of bits that can be encoded in
a channel that uses these mixtures. 
If the pure states $\ket{\psi_0}$ and $\ket{\psi_0'}$ are separated by a angle of $\theta$ on the surface of the Bloch sphere, 
then for small $\theta$, the number of distinct mixtures of two orthogonal pure states which can be distinguished at the $1\sigma$ level is approximately $8/\theta^{2}$.
% $\rho$ and $\rho'$
Thus, the maximum number of bits that can be encoded among these states can be
calculated as $\log_2{(8/\theta^{2})}$ \footnote{Our estimate of $8/\theta^{2}$ is an upper bound on the number of distinct mixtures that can be employed in this protocol. We note that the calculation of the number of states with a given angular separation that can be spread over the surface of a sphere is a non-trivial problem. However, for the angles we present in this work, no analytical solutions have  been found yet \cite{lai2023iterated}}. 
%Some work has been done in order to find the angle of separation between given numbers of points spread on the surface of a sphere.
For each of (a) to (d) in \Fig{multpairsplots}, we calculate the number of
bits to be 6.68, 8.04, 11.51, and 14.68, respectively. 

Reducing the angle by approximately a factor of 3 (from $10^{\circ}$ to $3^{\circ}$ in Fig.~\ref{fig:multpairsplots} (b) to (c)) increases the information per symbol by approximately $2\log_{2}{3}$, or 3.2 bits, but the number of photons required to resolve the symbols rises from $\sim 1.4 \times 10^3$ in (b) to $\sim 2.3 \times 10^4$ in (c).
From the four values of $\theta$ in Fig.~\ref{fig:multpairsplots} (a) to (d), we observe a general trend in the minimum number of photons required to distinguish the states separated by $\theta$, $N_{\mathrm{req}}$, scales roughly as $N_{\mathrm{req}} \sim 1/{\theta^2}$.
Thus, despite the increase in symbol bit capacity, the number of bits per photon drops from about $5 \times 10^{-2}$ to about $5 \times 10^{-3}$ --~the logarithmic scaling of the bit capacity suggests that a small symbol alphabet may be preferable in many communications settings.

\section{Conclusion}
We have demonstrated a way to learn the two single-photon polarization states
in statistical mixture from just the single-photon tomography data, provided that we have access to the pairing information, inspired by \cite{statelearning}. 
We used this data to `pair up' the photons based on their time-of-arrival
information and from here we obtain a 2-photon density matrix. 
Following \cite{statelearning}'s procedure, we uniquely decompose this
pair-state density matrix to find the two pure states and their probabilities.  
We perform an experiment to demonstrate this for different cases of
$\ket{\psi_{0}}$, $\ket{\psi_{1}}$, $p_{0}$ and $p_{1}$.
With on the order of $10^4$ detected pairs of photons we are able to learn the
two states to fidelities of around 99.99\%.
We also demonstrate how close two states in the same ensemble can be whilst we
can still discriminate between them.
For ${N \sim 10^4}$ pairs of photons, we can resolve two distinct states
$\ket{\psi_{0}}$ and $\ket{\psi_{1}}$ as close as $15^{\circ}$ to each other. 
We also see the closest two different mixtures can be to each other and how
many photons it takes to distinguish between the two. \\

\section*{Acknowledgments}

This work was supported by NSERC under Discovery Grant RGPIN-2020-05767, the QuEnSi quantum alliance (NSERC ALLRP 578468 - 22), and the John Templeton Foundation under grant ID 63209. Additional support came from the Fetzer Franklin Fund of the John E. Fetzer Memorial Trust. AMS is a fellow of CIFAR.
We sincerely thank for the helpful exchanges with An-Ning Zhang and his group
at the Beijing Institute of Technology, who are performing another
pair-learning experiment.
MH was supported by the National Science Foundation under the grant
Collaborative Research: NeTS: Medium 2504622.
BGE is extremely grateful for the long-standing support from the Centre for
Quantum Technologies, Singapore, where part of his share of the work was
done.

%%%%% one of the follwing two lines only, not both
% \bibliography{shortJournalNames,PairLearningExp.bib}

\bibliography{PairLearningExp.bib}

\end{document}